\documentclass[12pt]{iopart}
\pdfoutput=1
\usepackage{graphicx}
\usepackage{soul}
\usepackage{cite}
\usepackage{color}
\usepackage{float}

\usepackage{iopams}

\begin{document}

\title[Intrinsically disordered proteins]{Intrinsically Disordered Proteins at the Nano-scale}

\author{T. Ehm$^{1,2}$, H. Shinar$^1$, S. Meir$^1$, A. Sekhon$^1$, V. Sethi$^1$, I.~L.~Morgan$^3$, G. Rahamim$^1$, O.~A.~Saleh$^{3,4}$ and R. Beck$^1$}
\address{$^1$ The School of Physics and Astronomy, The Center for Nanoscience and Nanotechnology, and the Center for Physics and Chemistry of Living Systems, Tel Aviv University, Israel}
\address{$^2$ Faculty of Physics and Center for NanoScience, Ludwig-Maximilians-Universität, München, Germany}
\address{$^3$ BMSE Program, University of California, Santa Barbara CA 93110}
\address{$^4$ Materials Department, University of California, Santa Barbara CA 93110}

\ead{roy@tauex.tau.ac.il}

\begin{abstract}

The human proteome is enriched in proteins that do not fold into a stable 3D structure. These intrinsically disordered proteins (IDPs) spontaneously fluctuate between a large number of configurations in their native form. Remarkably, the disorder does not lead to dysfunction as with denatured folded proteins. In fact, unlike denatured proteins, recent evidences strongly suggest that multiple biological functions stem from such structural plasticity. Here, focusing on the nanoscopic length-scale, we review the latest advances in IDP research and discuss some of the future directions in this highly promising field.

\end{abstract}

%
\vspace{2pc}
\noindent{\it Keywords}: Intrinsically disordered proteins, nanoscopic characterization, single-molecule force spectroscopy, SAXS, FRET, NMR, FCS  
%
%
%
%

\section{Introduction}
The conventional paradigm in structural biology draws a direct connection between the amino-acid sequence of a protein to a singular 3D structure. The unique structure is considered essential to the protein's biological function by permitting only specific interactions. 

\begin{figure*}
\centering
\includegraphics[width=0.9\linewidth]{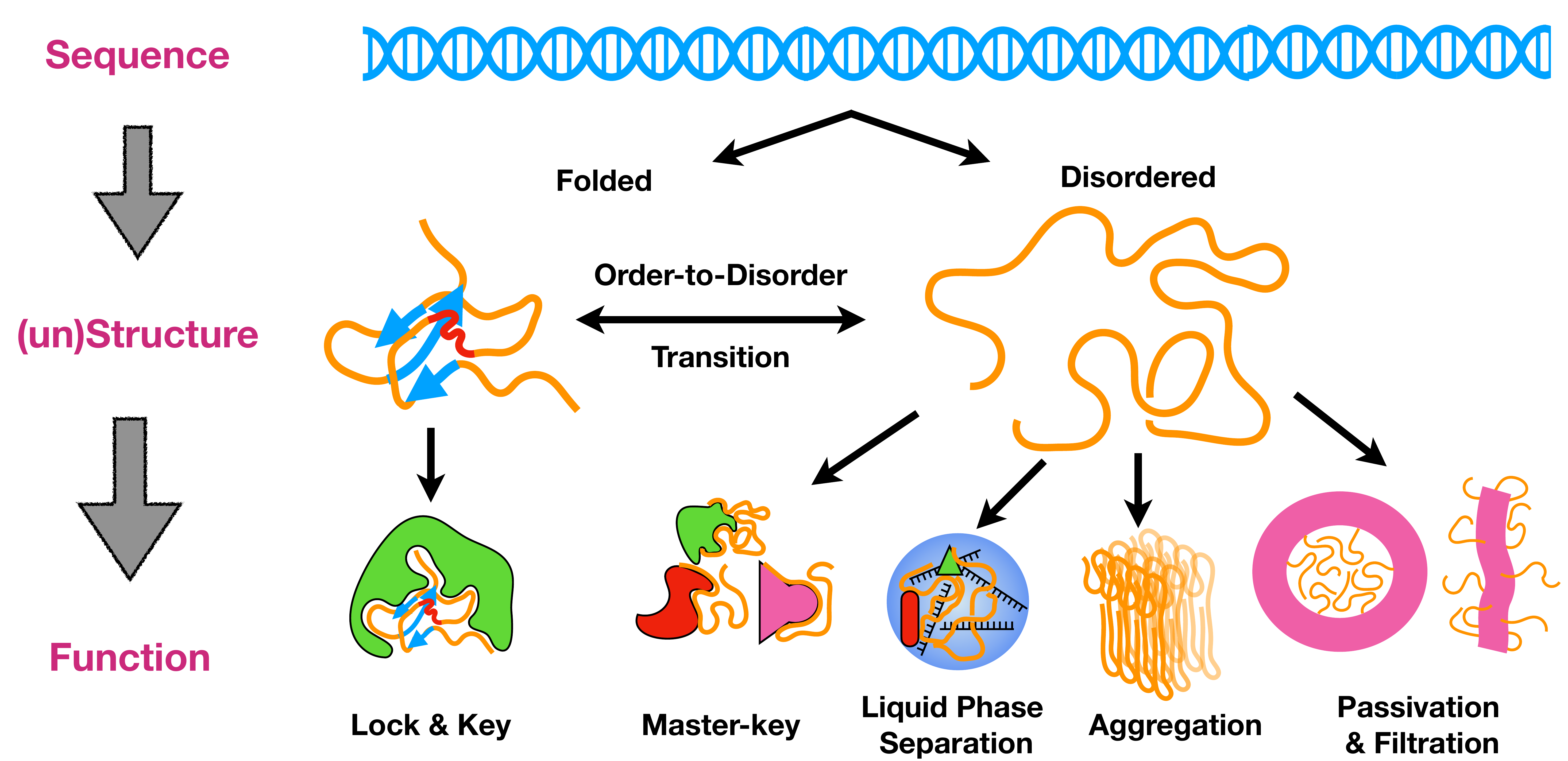}
\caption{
Contemporary sequence-function paradigm. The folded and disordered conformations, and the transition between the two, lead to biological function.}
\label{fig:structure-function}
\end{figure*}

 In the last two decades it has been recognized that up to 40-50\% of the proteome does not fit this simplified convention (Fig. \ref{fig:structure-function}). Instead, intrinsically disordered proteins and regions (IDP/IDR) fluctuate between a large number of conformations while still retaining their biological functions \cite{wright1999intrinsically,Uversky2014,dunker2001intrinsically,Uversky2010a,Uversky2011a}. An IDR is usually defined as an unstructured amino-acid stretch as part of a (folded) protein, and an IDP as a complete protein that does not fold to a stable 3D structure \cite{Uversky2011a,Uversky2010a,Uversky2014,dunker2001intrinsically}. For brevity, we will also use the IDP term to describe proteins having IDRs.

\begin{figure*}
\centering
\includegraphics[width=1.\linewidth]{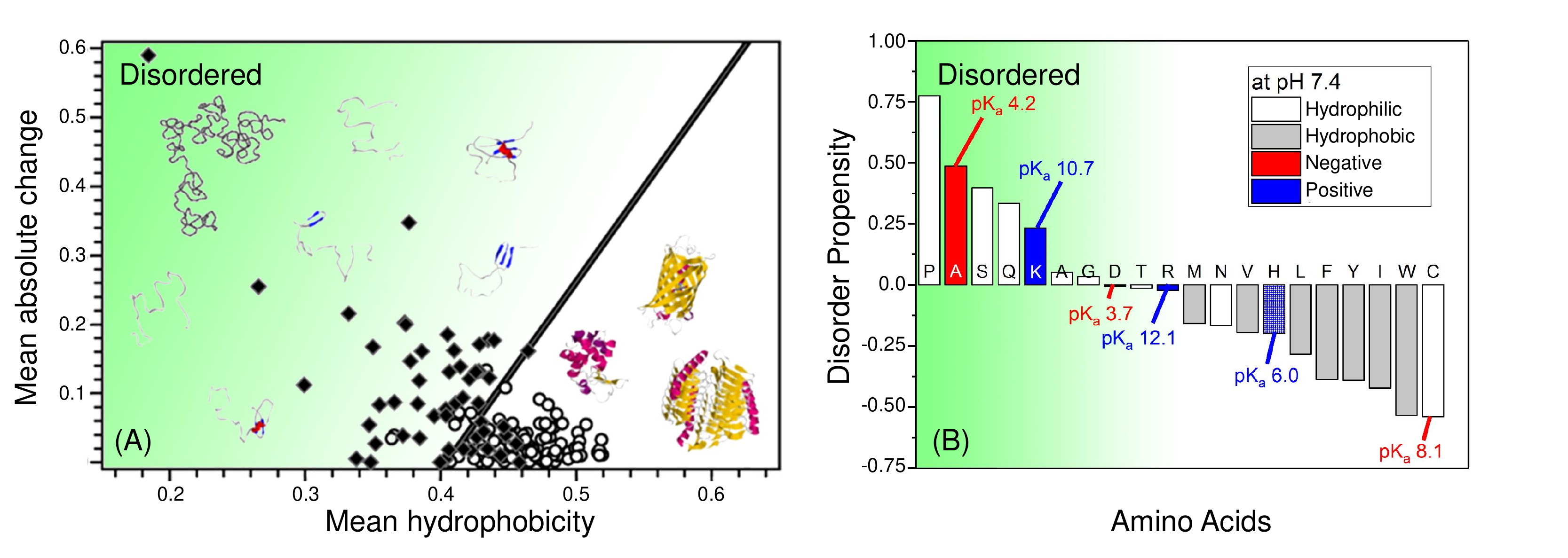}
\caption{The role of charge, hydrophobicity and amino acid residue in IDPs. (A) Nearly 250 folded (open circles) and nearly 90 natively unfolded proteins (black diamonds) demonstrate that IDPs are charged and hydrophilic. The solid line empirically separates between IDPs and compact globular proteins. Panel is adapted from \cite{Uversky2011a}. (B) The contribution of each amino acid in promoting  disorder. Disordered propensity is evaluated from the fractional difference of amino acids composition of IDPs in the DisProt database and a completely ordered proteins from the protein-database (PDB). Panel is adapted from \cite{radivojac2007intrinsic}.
}.
\label{fig:Charge and hydrophobicity}
\end{figure*}

Typically, most IDP sequences are rich in structure-breaking charged and polar amino acids, and depleted in order-promoting hydrophobic residues (Fig. \ref{fig:Charge and hydrophobicity}, and Refs. \cite{Kornreich2015a,Uversky2010a,Uversky2011a,Uversky2014,Uversky2020,uversky2000natively}). Indeed, computer-based methods exploit the amino-acid propensity and sequence as a sign of a disorder \cite{ferron2006practical}. Generally, IDPs fall into three distinct compositional classes that reflect the fraction of charged versus polar residues: polar, polyampholytes, and polyelectrolytes \cite{das2015relating,mao2013describing}. In addition, the balance between solvent mediated intra-chain attraction and repulsion directly influences the IDP's compactness. The compactness, in-turn, determines the accessibility to interact with other biomolecules. Therefore, modeling the inter- and intra-molecular IDP interactions, as well as the resulting ensemble of conformations, is an extremely complex and relevant problem in the research of nano-scale systems.

It is possible to characterize the IDP ensemble average parameters such as the end-to-end distance ($R_{ee}$) or the radius of gyration ($R_G$) distributions. In addition, the IDP length, charged amino-acid distribution, and propensity to form transient bridges, dictate the ensemble physical properties and can be connected to polymer physics theories. This could lead to new language that relates the biological function to basic physical parameters. Thus, IDP research has the opportunity to gain a novel insight into of the biological sequence-to-function problem (Fig. \ref{fig:structure-function}).

\section{Biological function}  

The structural plasticity of IDPs is the key to their function. This plasticity, almost by definition, limits the strength of the IDPs' interactions with other biomolecules and the environment. Interactions larger than the thermal energy would ultimately lead to a specific configuration, which conflicts with the IDPs' large ensemble of conformations \cite{chakraborty2019nanoparticle,shin2017liquid}. Therefore, we must ask ourselves: does structural plasticity compete with specificity?

Distinct from many structural proteins, IDPs can interact with multiple different partners. In analogy to the lock-and-key concept for structural proteins, IDPs serve as ``master-keys'', each capable of openning many different ``locks'' \cite{wright1999intrinsically,dunker2001intrinsically,Uversky2020}. However, given IDP's prevalence, several other functions have been reported including cellular signaling pathways, a regulator for protein-protein and protein-DNA networks or folding into ordered structures upon binding to other cellular counterparts, to name a few (Fig. \ref{fig:structure-function}) \cite{wright2015intrinsically,marmor2020spatio,dunker2005flexible,gianni2016coupled}. 

In several cases disorder-to-order transition of IDPs is the ensemble's conformational response to various stimuli. Such response can lead to collapse or expansion of the IDP, e.g. globule-to-coil or collapse transitions \cite{das2015relating,Uversky2020,kodera2020structural}. Recently, high-speed AFM measurements demonstrated constantly folded and disordered regions as well as disorder-to-order transitions in IDPs \cite{kodera2020structural}.

Unraveling the factors governing the IDPs' ensemble conformation and their disorder-to-order transition is  crucial to understand their biological functionalities \cite{zhou2019intrinsically}. An illustrative example is the regulation of glucose homeostasis by human pancreatic glucokinase (GCK) enzyme. GCK catalyzes glucose catabolism in the pancreas. At low glucose concentration, IDR of GCK associates with glucose and undergoes a disordered-to-ordered transition. Following glucose release, the IDR undergoes order-to-disorder transition on millisecond time-scale. This “time delay loop” allows slow turnover and kinetic cooperativity of the enzyme. At high glucose concentration, the delay loop is bypassed and excess glucose is catabolized \cite{larion2012order}.

\begin{figure*}
\centering
\includegraphics[width=1.\linewidth]{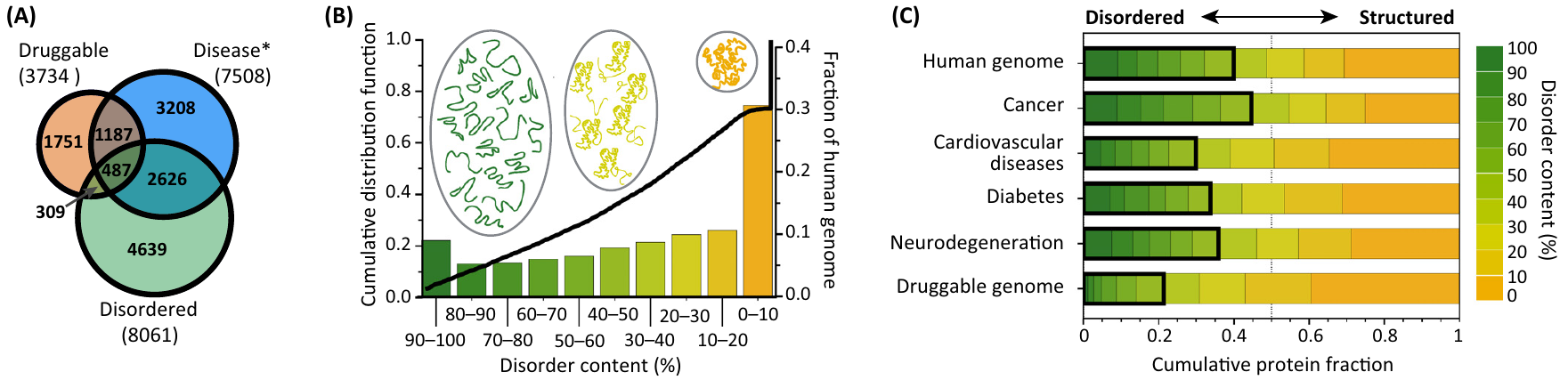}
\caption{
Disordered proteins links to human diseases. (A) Venn diagram of three types of proteins: Those that interact with drugs; those that are related to disease; and those that are disordered. (B) Fraction of humans proteins (right axis) binned according to their content of structural disorder. Black line is the cumulative distribution function (left axis). (C) Amount of disorder in different protein categories. Figure adapted from Ref. \cite{heller2015targeting}. }
\label{fig:IDP-prevalence}
\end{figure*}

Obviously, the function of IDP is driven by their amino-acid sequence with mutations resulting in numerous diseases (Fig. \ref{fig:IDP-prevalence}). Known examples are the disordered regions found in the Alzheimer and Parkinson associated proteins Amyloid-$\beta$ and $\alpha$-synuclein, that can form toxic oligomers, amyloid fibrils, and other types of aggregates \cite{carballo2017comparison, coskuner2018insights}. Other diseases, such as the recently emerged novel coronavirus (SARS-CoV-19), are intimately related to protein disorder. The SARS-CoV-19 genome sequences encode an IDR in its nucleocapsid, an essential structural component that binds to RNA and interacts with several proteins in its multitasking role \cite{ceraolo2020genomic,cubuk2020sars}. 

IDPs are also present in many biological receptors and play a functional role which is often poorly understood \cite{richardson2009acidic, chino2019intrinsically}. For example, the role of the disordered acidic domains of the Toc159 chloroplast preprotein, which binds to transit peptides, is still unknown \cite{richardson2009acidic}. Many RNA binding proteins (RBP) also contain disordered regions. There, the disordered domains play a central role in regulatory processes (e.g., stability) \cite{basu2016structural}. 
One fascinating class of IDPs comprises neuronal-specific cytoskeleton proteins, including neurofilament (NF) proteins, $\alpha$-internexin, vimentin, microtubule-associated protein 2 (MAP2), and tau \cite{Pregent2015,Kornreich2015a,Laser-Azogui2015,Safinya2013,Guharoy2013,Chung2016}. A key motif displayed by these proteins involves long IDRs mediating the assembly of filaments into a hydrogel network \cite{Chung2016,Kornreich2015a,Kornreich2016}. This network is responsible, for example, for defining the structure, size, and mechanics of axons, with direct effects on electrical conduction \cite{Laser-Azogui2015,Chang2004,Malka-Gibor2017,Beck2010,Chung2016}. Interestingly, the basic structural properties of these networks can be understood using the polymer-physics theoretical arsenal. Nonetheless, the full functional behaviour is sequence specific, and so can not be coarse-grained \cite{Kornreich2015a,Malka-Gibor2017,Chung2016}.     

To cope with the enormous traffic of molecular exchange across the nuclear envelope, nuclear pore complexes (NPCs) have evolved to exchange cargo rapidly and with high selectivity. NPCs are composed of ~30 different nucleoporins that assemble into sub-complexes to build the NPC's distinct modules. Though it is not entirely clear how the cytoplasmic filaments are anchored to the NPC core, the central, disordered protein segments of Nup145N, Nup100 and Nup116 are known integral parts of this interaction \cite{Schwartz2016,Lemkeshi2016}.

Recently, much attention has been drawn to membraneless organelles, which are  liquid-like condensates formed reversibly by dynamic self-assembly of proteins, mostly IDPs, and RNA through a liquid-liquid phase separation (LLPS) \cite{shin2017liquid,Brangwynne2009,marmor2020spatio}. This behavior has raised many questions: Does the high density within such condensates enable certain protein functions? Is it the proteins' way of defending themselves against degradation? Or is LLPS just an unwanted outcome of the IDPs' sequence, which leads to transient interactions with multiple partners? The answers to this questions are most likely different for alternative organelles compositions.

 There is little understanding of the IDP's sequence-encoded mechanism that drives this phase separation. Regardless, membraneless organelles constituents often contain multiple repetitive sequences, facilitating multivalent, weak interactions with their partners to form the condensates. As mentioned, such interactions are frequent for IDPs. The size of membranelss organelles are in the range of sub $\mu$m to 10 $\mu$m and thus can be observed with optical microscopy. For example, the disordered regions of Ddx4 \cite{Nott2015}, LAF-1 \cite{Elbaum-Garfinkle2015}, FUS \cite{Patel2015}, which are rich in glycine, form droplets of ~1$\mu$m.
 
 Even folded proteins that have IDRs, such as hnRNPA1 \cite{Molliex2015} and TDP-43 \cite{Conicella2016}, form droplets, designated as stress granules \cite{buchan2009eukaryotic}. Assembly and disassembly of these granules is a highly regulated process involving the IDR interactions with multiple proteins and mRNAs. Recently, using multi-bait engineered ascorbate peroxidase (APEX) proximity labeling technique, over a hundred proteins were discovered that either promote assembly or induced disassembly of stress granules \cite{marmor2020spatio}. Importantly, APEX revealed that most of these proteins are indeed disordered. Such studies are extremely relevant for in-depth understanding of the compositional changes in stress granules' proteins in neurodegenerative disorders.

\section{Artificial IDPs and polymer physics}

Disordered proteins' ability to engage in a variety of manners often results in a rich ensemble of phase transition behaviors \cite{dignon2018sequence,zhou2018disordered}. These phase behaviors can be easily manipulated by slightly changing the physio-chemical properties of the IDPs. Recently, the Chilkoti group studied a simplified version of artificial IDPs, all made of a repetitive 8-mer peptide sequence originated from Drosophilia melanogaster Rec-1 resilin \cite{dzuricky2020novo}. Interestingly, changing a single amino acid in the repeat, or increasing the number of repeats, results in a drastic and robust change in the LLPS temperature. The IDP sequence also controls the temperature-ramp hysteresis phenomena. Notably, this hysteresis changes when reversing the amino acid sequence originating from N- to C-terminal \cite{quiroz2019intrinsically}. These findings demonstrate the ability to fine-tune the phase behaviors and indicates to the significance of IDP's sequence, as in structural proteins. 
\begin{figure}
\centering
\includegraphics[width=1.\linewidth]{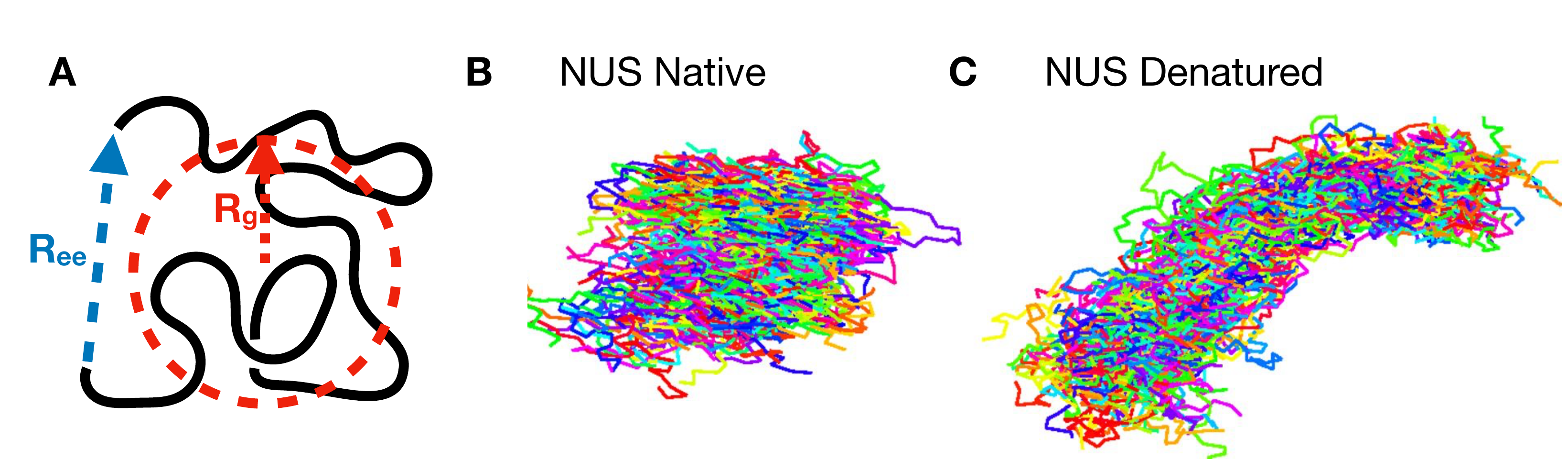}
\caption{
(A) Ensemble structural characterization of IDP that includes the end to end distance ($R_{ee}$), typically measured by FRET, and radius of gyration ($R_{g}$), measured by SAXS. Representative ensembles for NUS protein in its (B) native condition and (C) in denatured condition. Both ensembles recapture FRET and SAXS data at these conditions and show larger asphericity in denatured conditions. Panels (B) and (C) are adapted from Ref. \cite{fuertes2017decoupling}. }
\label{fig:schmatics_and_NUS}
\end{figure}

Given IDPs' structural plasticity, it is tempting to investigate them as polymers, at least to first order (Fig. \ref{fig:schmatics_and_NUS}). By doing so, the power of statistical-mechanics can be harnessed to quantify the volumetric dimension of IDPs. Then, coarse-grained structural descriptors are obtained from the scaling laws of polymer physics \cite{wang201750th}. Specifically, the degree of IDP collapse or expansion can be quantified in terms of chain's length and its relation to the ensemble's radius of gyration: $R_g \propto N^{\nu}$. Here, $N$ is the number of monomers, and $\nu$ is the Flory scaling exponent \cite{wang201750th,Kornreich2015a}.

Qualitatively, $\nu$ determines the balance between intra-chain and solvent-chain interactions, also referred to as solvent quality. For good ($\nu\approx 0.6$), $\theta$ ($\nu$=0.5), or poor ($\nu \approx 0.3$) solvents the conditions are respectively less, equal or favorable intra-chain interactions respective to solvent/chain ones. Thus, the value of the scaling exponent $\nu$ can act as a physical descriptor for the collapse transition as a function of different environmental conditions \cite{riback2017innovative}. In the following, we will introduce several techniques targeting the nanoscopic dimensions of IDPs.

\section{Nanoscopic techniques}

The detection and characterization of disordered proteins and regions are of great interest in order to determine their function. IDPs' structural plasticity dictate techniques and analyses targeting representative parameters of the protein's conformational ensemble \cite{LazarPED2021}. Naturally, we will not be able to discuss the entire breadth of techniques; instead, we will focus on some of those targeting the nano-scale regime, which is, in most cases, the relevant length-scale. 

\subsection{Simulations}
To further understand the conformational dynamics of IDPs, advanced molecular dynamics (MD) simulations are used. However, the inherently large number of degrees-of-freedom, the inaccuracies in the simulation models, and the need for long simulations have stymied progress. To overcome these limitations, alternative approaches have been developed. For example, a hierarchical approach \cite{pietrek2019hierarchical} based on the “flexible-meccano” model \cite{bernado2005defining} is generating representative and meaningful starting ensembles. Alternatively, in the "sample-and-select" approach, sub ensembles are derived from a broader distribution of molecular conformations based on experimental NMR and small-angle X-ray scattering (SAXS) data \cite{jensen2014exploring, fisher2011constructing, mittag2007atomic, allison2009determination}. 

A recent computational work \cite{baul2019sequence} developed a related strategy: experimental SAXS data on a few select IDPs was used, through a fitting approach, to develop a set of sequence-specific residue interaction parameters in a coarse-grained simulation. These parameters were then used to investigate the conformational ensembles of a range of IDPs. The authors found the IDPs displayed apparent polymeric (swollen chain, $\nu=0.6$) behavior when considering $R_g$ vs. $N$. Yet, the simulated structural ensembles showed significant heterogeneity: certain IDPs showed elevated intra-segment contacts and transient sub-segment compactions. However, this was not universal: other IDPs in the study behaved as homogeneous random-walk chains. The authors concluded that sequence-specific structure could be hidden behind polymeric exponents\cite{baul2019sequence}.

\subsection{SAXS} 
A common and powerful technique suited for IDP structural characterization is SAXS. The technique's key advantage is measuring the ensemble conformations while kept in solution without the need for crystallization or external tagging. A simple SAXS analysis immediately provides $R_g$ and the pair distance distribution \cite{kornreich2013modern,manalastas54atsas}. Additionally, traditional Kratky plots ($q^2I(q)$ vs. $q$), directly obtained from the SAXS profiles of IDPs, provide a qualitative picture of the presence of globular or unfolded conformations \cite{bernado2012structural, almagor2013structural}. Here, $I$ is the scattering intensity, and $q$ is the scattered wave-vector proportional to the scattering angle.  

Moreover, the scattering profile can be used to extract the Flory exponent ($\nu$) mentioned above \cite{hammouda2012small, fuertes2017decoupling}. More advanced analysis techniques are regularly used to extract the ensemble's dominant conformations from the SAXS signal (Fig. \ref{fig:schmatics_and_NUS}). The ensemble optimization method \cite{bernado2007structural, tria2015advanced}, and molecular form-factor \cite{riback2017innovative} 
are just a few examples for such techniques resulting in representative conformations that fit the experimental data. We note that the combination of SAXS with other complementary methods (Fig. \ref{fig:schmatics_and_NUS}) such as simulations, NMR, and F{\"o}rster resonance energy transfer (FRET) provides a more comprehensive picture of IDPs, and their conformations in solution \cite{fuertes2017decoupling,storm2015liquid, LazarPED2021,Malka-Gibor2017,kornreich2013modern,Kornreich2015a,storm2016loss}. 

\subsection{FRET}  
The distance-dependent dipole-dipole interactions between probes bound to the side chains of IDPs provide the basis for determining long-range intra-molecular distances between selected sites. Experiments based on single-molecule \cite{Hofmann2012} or ensemble \cite{Grupi2011} measurements are characterized by $1-10$ nm distance resolution range and an ability to recover distributions of intra-molecular distances in the transient ensembles of IDP. For example, it was shown that $\alpha$-synuclein deviates from ideal chain behavior at segments labeled in the NAC domain and N-termini using the ensemble method. It was suggested that those conformations bias might be related to the initiation of amyloid transition \cite{Grupi2011}. Another example, using a single-molecule method, showed that the previously mentioned scaling exponents ($\nu$) in water strongly depend on the sequence composition. Two of the total examined IDPs did not reach the $\theta$-point under any solvent conditions. This may reflect their biological functional need for an expanded state optimized for interactions with cellular partners \cite{Hofmann2012}.

\subsection {NMR}
NMR spectroscopy offers a unique platform in deciphering IDP dynamics and structure \cite{konrat2014nmr}. The traditional NMR hydrogen chemical shift signal is, in many cases, insufficient for IDP characterization. Fortunately, new tools have been developed to overcome low signal to noise ratio difficulties such as paramagnetic relaxation enhancement (PRE) \cite{otting2010protein} and proline based 2D H-N residue correlations \cite{solyom2013best}. 

In PRE, the introduction of paramagnetic spin labels in proteins affects the chemical shift and transverse relaxation rate signal between the unpaired electron and NMR active nuclei on the basis of distance between them \cite{otting2010protein}. For example, the PRE signal and $^{15}N$ relaxation data was analyzed to quantify the interaction between the IDP osteopontin and heparin \cite{platzer2011metastasis}. There, it was found that on binding with heparin, osteopontin largely remains in a disordered state and undergoes structural/dynamical adaption which is mainly mediated by electrostatic interactions.  

A recent development using in-cell NMR spectroscopy provides the opportunity to explore the structural plasticity of an IDP in its native environment \cite{luchinat2016unique,barbieri2016characterization}. Both eukaryotic and prokaryotic IDPs have been investigated using in-cell NMR such as  $\alpha$-synuclein, prokaryotic ubiquitin-like protein, Pup, tubulin-related neuronal protein, tau, FG-Nups in nucleoporins, and the negative regulator of flagellin synthesis, FlgM. These studies revealed that the cellular conformational dynamics may differ significantly from these observed in-vitro \cite{sciolino2019cell}.

\subsection{FCS} 
Due to diffusion and the protein structural plasticity, the emission of fluorescently-labeled molecules fluctuate. In fluorescence correlation spectroscopy (FCS), these fluctuations are recorded within an illuminated confocal volume \cite{magde1972thermodynamic,elson1974fluorescence,elson2011fluorescence}. The timescale for conformational fluctuation (i.e., chemical kinetics within the biomolecule) lies in the nano- to microseconds range, whereas for translational diffusion from microsecond to milliseconds. Information about the diffusion coefficient and chemical kinetics can be inferred from the fluorescence fluctuations autocorrelation and cross-correlation signal. Furthermore, from the diffusion coefficient, it is possible to determine the molecular size and hydrodynamic radius of the investigated biomolecules. 

FCS is also sensitive to fluorophore quenching reactions. These measurements can provide information about the internal dynamics of the IDPs \cite{mukhopadhyay2007natively}. Fluorescence self-quenching of tetramethyl rhodamine, which is chemically labelled at two different residual positions, was analyzed to study the conformational kinetics of unfolded intestinal fatty acid binding protein \cite{chattopadhyay2005kinetics} and amyloid forming yeast prion protein, Sup35 \cite{mukhopadhyay2007natively}.

\begin{figure}
\centering
\includegraphics[width=1.\linewidth]{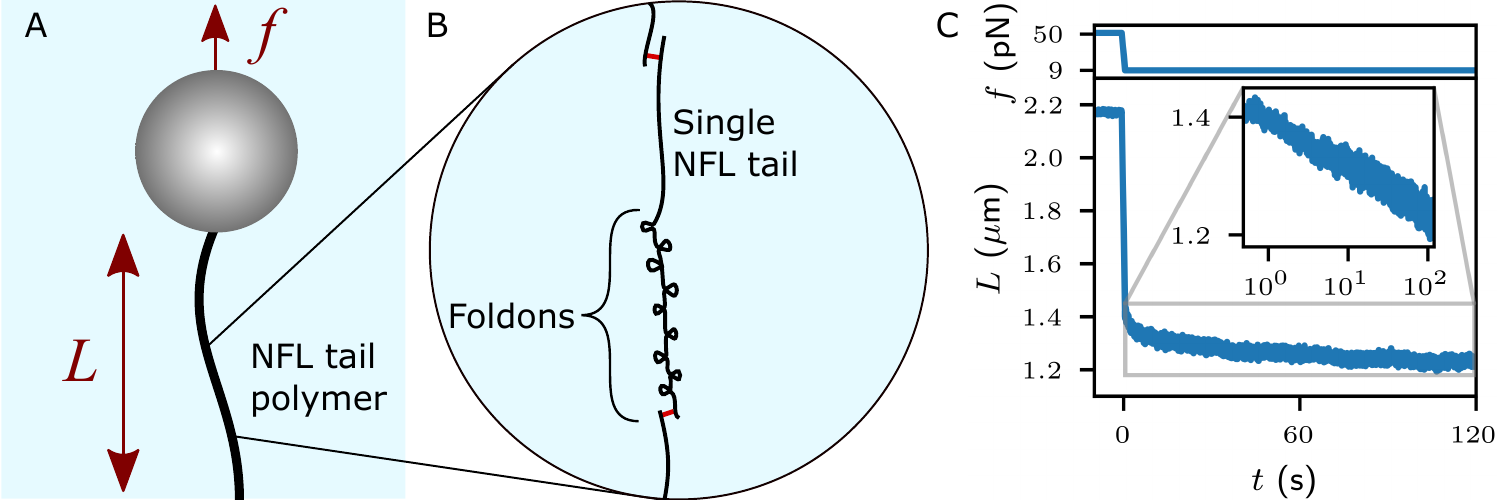}
\caption{
Mechanical perturbation reveals subtle internal IDP structure. (A) Under constant tension $f$, the length $L$ of a polymer of NFL tails is measured. (B) A single NFL tail contains multiple, independent regions with internal structure (foldons). (C) An initial large tension breaks apart internal structure in the NFL tails. After rapidly lowering the tension, at constant tension, the NFL polymer shows a logarithmic decrease in length over time $t$, indicating multiple regions of internal structure are reforming. Panel (C) is adapted from Ref.~\cite{morgan2020glassy}.}
\label{fig:nfl-smfs}
\end{figure}

\subsection{Single-molecule force spectroscopy}
Single-molecule force spectroscopy (SMFS) uses the thermodynamic effects of applied tension to gain  insight into the nanoscopic conformation of IDPs. SMFS consists of tethering an IDP between a static surface and a force probe and measuring the chain extension, with nanometer accuracy, as a function of force. This nanomechanical assay is  a high-precision differential measurement of the effect of varying perturbations (tensions) on IDP interactions and conformations. For example, AFM-based  experiments on several amyloid precursor IDPs show a sawtooth pattern that indicates the mechanical unfolding of multiple different conformations \cite{hervas2012}. Similarly, optical tweezer experiments on $\alpha$-synuclein show it has several marginally stable and rapidly fluctuating subsegments \cite{solanki2014}. 

A noteworthy recent work demonstrated the power of SMFS-based perturbation in revealing more subtle IDP structural behaviors \cite{morgan2020glassy}. In the work, a polyprotein of the disordered Neurofilament light-chain tail (NFLt) domain was subjected to force-quench experiments, in which a large tension pulled the chain straight, and was followed by a sudden jump to a small tension that permitted structure in the IDP to re-form (Fig.~\ref{fig:nfl-smfs}). The experiment was carried out in a magnetic tweezer, an SMFS instrument that offers high-stability in applied force, allowing the tracking of NFLt dynamics over long periods of time. The experiment revealed a glassy, non-exponential relaxation in which the chain extension decreased logarithmically with time for many minutes. This behavior, extraordinary for its slow speed, was attributed to multiple, small collapse events within a single NFLt domain, with each collapse changing chain extension by only $\approx 1$ nm.
This experiment highlights both the high sensitivity of SMFS, as well as the rich and physically-complex structural dynamics of IDPs.

\subsection{Quantifying IDP's interactions}

Intermolecular interactions of  biomolecules play an essential role in the proper function and growth of biological cells\cite{leckband2001intermolecular,wyttenbach2007intermolecular,musiani2017protein}. These interactions can be specific or non-specific.
Non-specific interactions are transient and weak, mainly governed by steric repulsion or ionic bridging. IDPs' function is largely dominated by these weak interactions. This was demonstrated, for example, in a recent AFM-based SMFS study, in which the interactions between nuclear transport factors and disordered FG repeats was found to display unexpected complexities due to weak, multivalent contacts between the interacting partners \cite{Hayama2019}.


Improved protein-protein interaction resolution can be achieved using bulk, low-throughput spectroscopic methods such as FRET, surface plasmon resonance (SPR), and nuclear magnetic resonance (NMR), to name only a few. For example, the interaction between the disordered region in transcription factors Sp1 and TAF4 was estimated by NMR to be $K_d =69 \mu$M \cite{Hibino2016}. IDP Ntr2 modulates the RNA Helicase Brr2 with $K_d = 14 \mu$M and $7\mu$M determined by SPR and NMR, respectively \cite{Wollenhaupt2018}. SPR was also used to detect and characterize IDP–ligand interactions with tau protein as a model IDP \cite{Vagrys2020}.

Another recently introduced technique aiming to overcome the technical difficulties to measure weak and transient protein-protein interaction is the nanoparticle mobility assay  \cite{chakraborty2019nanoparticle}. 
There, nanoparticles grafted with IDPs were imaged while diffusing over a surface grafted with a second set of IDPs. A similar approach was used to study strong DNA-DNA interactions \cite{Xu2011}, delocalized long-range polymer-surface interactions \cite{Skaug2014} and bulk-mediated diffusion on supported lipid-bilayers \cite{Yoo2016}. 
By carefully analyzing the particles' diffusive nature, the authors detected an altered interaction caused by a single mutation on the polypeptide sequence \cite{chakraborty2019nanoparticle}.

\section{Summary and perspective}

Two decades of IDP research place them at the forefront of proteomics\cite{wright1999intrinsically}.  Apparently, structural plasticity is valuable primarily because it enables the IDPs to adjust to their environment. In addition, IDPs are also used in various nano-biomedical technologies. For example, \emph{PASylation} technology involves conjugating pharmacologically active compounds with natively disordered biosynthetic polymers made of the small L-amino acids Proline, Alanine and/or Serine \cite{binder2017pasylation}. This technology may serve as an alternative to PEGylation, which is currently widely used to sterically stabilized therapeutic proteins and peptides but is often accompanied by immunogenicity and lack of biodegradability \cite{gebauer2018prospects, schlapschy2013pasylation}. IDPs can also serve to modify the  adhesion properties of surfaces. For example, a study on barnacle adhesive IDPs on silica surfaces showed modified protein-surface affinities, adhesive activities and kinetic adsorption rates \cite{wang2018adsorption}.

IDPs' large repertoire has been implicated in numerous diseases, making them potential targets for therapeutic intervention. Disease-associated IDPs can serve as potential targets for drugs modulating protein-protein interaction networks that participate in both the ``one to many'' and ``many to one'' interaction \cite{wang2011novel}.

The relation between IDPs and polymer- and statistical-physics is continuously revisited. Na{\"\i}vely, IDP \emph{function} could have been understood from coarse metrics such as the IDP's length, total charge, net hydrophobicity, etc. However, as with folded proteins, the sequence of amino acids, and not just the net composition, does dictate the resulting function, at least to some extent. Yet building a coarse-grained model for IDPs is a highly complicated and fascinating problem due to their unique properties. The most fundamental difficulty is somewhat technical - the large ensemble of IDP conformations is often impossible to sample with modern computers \cite{csizmok2016dynamic}. Another problem evolves from the extended amount of electrostatic interactions that frequently exist in IDP systems. Those interactions tend to span over a broad range of length and time scales, and hence are hard to quantify by a finite force field \cite{cragnell2018utilizing}, although new IDP-specific force fields have been suggested  \cite{knott2014discriminating,sigalov2016structural,zhao2020investigating,schuler2016single}. Nevertheless, further research is highly needed, especially since many phenomena involving IDPs are still not fully understood \cite{bhattacharya2019recent}.     

\ack

This work was supported by the National Science Foundation under Grant No. 1715627, the United States-Israel Bi-national Science Foundation under Grant No. 2016696, and the Israel Science Foundation under Grants No. 1454/19.

\section*{References}
\bibliographystyle{iopart-num}
\bibliography{lib}

\end{document}